\documentstyle[twoside,fleqn,espcrc2,epsf]{article}

\newcommand{\AmS}{{\protect\the\textfont2
  A\kern-.1667em\lower.5ex\hbox{M}\kern-.125emS}}

\newcommand{\bee}{\begin{equation}}
\newcommand{\ee}{\end{equation}}
\newcommand{\beea}{\begin{eqnarray}}
\newcommand{\eea}{\end{eqnarray}}

\title{Kaon B-parameter using Overlap Fermions}

\author{Thomas DeGrand (MILC collaboration)\\[2mm]
        Department of Physics, 
        University of Colorado\\ Boulder, CO 80309-390, USA}

\begin{document}

\begin{abstract}
I present first results from an in-progress
 calculation of $B_K$ in quenched approximation using overlap
fermions.
My particular implementation of the overlap uses a kernel with nearest
and next-nearest neighbor interactions and HYP-blocked gauge connections.
Matching to the continuum NDR regularization is done perturbatively.
I present preliminary results at $\beta=5.9$ and 6.1 (lattice spacings 0.125
and 0.09 fm) for quark masses, pseudoscalar decay constants, and 
B-parameter -- $B_K^{(NDR)}(\mu=2$ GeV) $\simeq 0.66(3-4)$.
\end{abstract}

\maketitle
The kaon B-parameter $B_K$,  defined as
${8\over 3} (m_K f_K)^2 B_K =\langle \bar K| \bar s \gamma_\mu(1-\gamma_5) d
\bar s \gamma_\mu(1-\gamma_5) d | K \rangle $,
has been computed many times with lattice methods.
Lattice calculations of $B_K$ require actions with good chiral properties,
to prevent operator mixing with wrong-chirality operators from contaminating
the signal.
 There has been a continuous  cycle of lattice calculations using
 fermions with ever better chiral properties.
This calculation is yet another incremental upgrade, to  the use
of a lattice action with exact $SU(N_f)\otimes SU(N_f)$ chiral symmetry,
an overlap action. These actions have operator mixing identical to that
of continuum-regulated QCD.

The overlap action used in these studies\cite{ref:TOM_OVER}
 is built from a kernel action with nearest and
next-nearest neighbor couplings,
and  HYP-blocked links\cite{ref:HYP}.
HYP links fatten the gauge links
without extending gauge-field-fermion couplings beyond a single hypercube.
This improves the kernel's chiral properties without compromising locality.
The kernel action is designed to resemble the exact overlap well enough
that its eigenvectors are good ``seeds'' for a calculation of
eigenvectors of the exact action, and it is kept simple enough that
finding its own eigenvectors is inexpensive.
These eigenvectors 
are used to precondition the calculation of
quark propagators, in principle eliminating {\it all} critical
slowing down at small quark mass. 

The data set is generated in the quenched approximation
 using the Wilson gauge action at  couplings
 $\beta=5.9$ (on a $12^3 \times 36$ site lattice) 
and $\beta=6.1$ (on a $16^3 \times 48$ site lattice) 
with 40  lattices each (so far).
 The nominal lattice spacings are $a=0.125$ fm and 0.090 fm 
from the measured rho mass.
Propagators for six quark masses are constructed
corresponding to pseudoscalar-to-vector meson mass ratios
of $m_{PS}/m_V \simeq0.6$ to 0.85.

\begin{table}[htb]
\caption{Results from these simulations.}
\label{table:1}
\vspace{0.05in}
\begin{center}
\begin{tabular}{lcc}
\hline
    & $\beta=5.9$ & $\beta=6.1$ \\
\hline
$1/a$ (MeV)  & 1580(60)  & 2190(140) \\
$m_{nonstrange}$ (MeV) & 4.3(3) & 4.5(3) \\
$m_{strange}$   (MeV) & 105(5) & 110(7) \\
$m_s/m_{ns}$ & 24.40(4) & 24.41(5) \\
$f_\pi$ (MeV) & 142(11)  & 131(12)  \\
$f_K$ (MeV) & 155(10) & 147(11) \\
$B_K^{(NDR)}(\mu=2$ GeV) & 0.66(3) &0.66(4) \\
$B_K^{(RGI)}$ & 0.92(4) &0.93(6) \\
\hline
\end{tabular}
\end{center}
\vskip -1.2truecm
\end{table}

The methodology for $B_K$ is well-developed:
compute an un-amputated correlator which contains the desired 
matrix element (two kaon sources
 far apart on the lattice with the four-fermion 
operator sandwiched
in between), clip off the $(m_K f_K)^2$ prefactor by simultaneously
 measuring the matrix element
 $\langle 0 | \bar s \gamma_0 \gamma_5 d | K\rangle$,
extrapolate/interpolate the lattice $B-$parameter to
 its value at  the kaon mass, and convert
the lattice number to its continuum-regulated counterpart.

To maximize the  signal volume I computed propagators from two
 well-separated sources ($N_t/2 -2$ temporal sites apart) and brought them
together to the operator.
I used Gaussian sources to maximize overlap onto the ground state.
These sources do not make  momentum eigenstates, and so
the $p=0$ $B_K$ signal is contaminated by a $\vec p \ne 0$ contribution.
This causes problems at bigger quark mass, because $E(p)-m$ gets
smaller as the pseudoscalar mass $m$ grows.
 Fortunately, there are two inequivalent
 paths on the torus to disentangle
the two ``signals,'' and one can fit the $B_K$ correlator to a sum
of a $\vec p=0$ term and a $p=2\pi/N_s$ term.
This is not a problem at small quark mass.

\begin{figure}[h!tb]
\begin{center}
\leavevmode
\epsfxsize=70mm
\epsffile{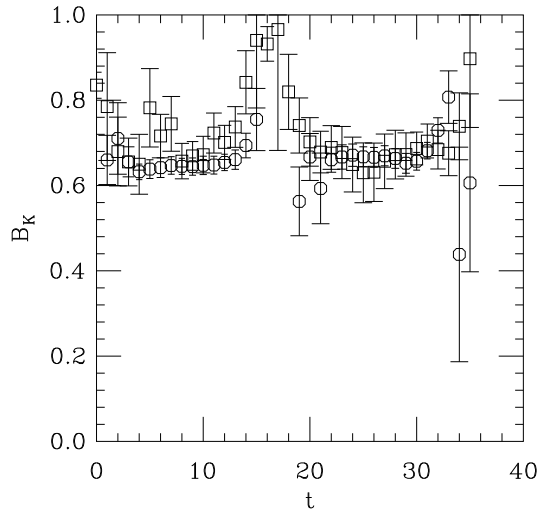}
\end{center}
\vspace{-28pt}
\caption{
A traditional ratio plot of the $B_K$ graph divided by the product of
two point graphs, from the $\beta=5.9$ data set at quark mass $am_q=0.050$
with axial current sources  and sinks
(squares) and pseudoscalar sources and sinks (octagons).
}
\label{fig:bk0.050rat}
\end{figure}

\begin{figure}[htb]
\begin{center}
\leavevmode
\epsfxsize=70mm
\epsffile{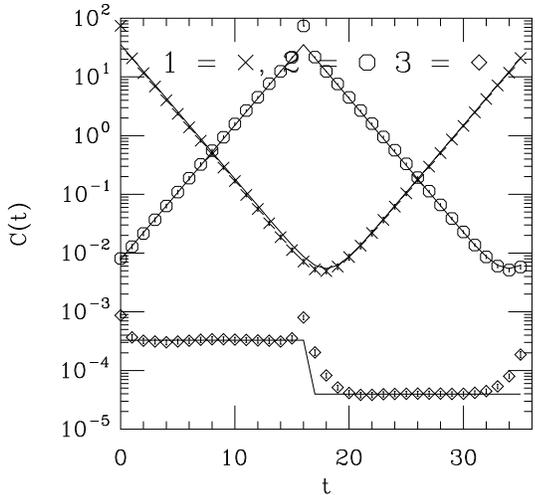}
\end{center}
\vspace{-28pt}
\caption{The two axial current correlators (labeled ``1'' and ``2'') and the
``figure-eight'' correlator (labeled ``3''), for the $am_q=0.100$ $\beta=5.9$
data set with axial current sources. A correlated fit to three correlators
over the range $t=7-9$ and 24-28 is also shown.
}
\label{fig:bkax0.100}
\end{figure}

\begin{figure}[h!tb]
\begin{center}
\leavevmode
\epsfxsize=70mm
\epsffile{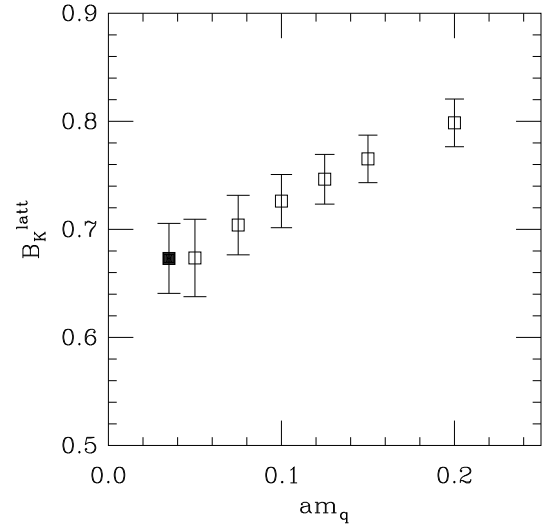}
\end{center}
\vspace{-28pt}
\caption{
Lattice $B_K$ (from ratio fits) 
at $\beta=5.9$, with a jackknifed linear extrapolation to the kaon (half
the strange quark mass) (solid symbol).
}
\label{fig:jbka}
\end{figure}

I extracted a signal from fits to the traditional ratio of
the $\bar K - K$ amplitude and product of two-point functions, as well as
correlated fits to the $\bar K - K$ amplitude
and the two-point functions. 
 A   ratio plot
is shown in Fig. \ref{fig:bk0.050rat}.
A typical  correlated fit is shown in Fig. \ref{fig:bkax0.100}, and a
plot of lattice $B_K$ vs. quark mass is shown in Fig. \ref{fig:jbka}.
Uncorrelated jackknife ``ratio fits'' have small
uncertainties and are quite stable
over a wide range of timeslices. However, one would really like
to do correlated fits. The data points are strongly correlated,
and it is necessary to use singular value decomposition to invert
the correlation matrix. When I do this, I find that my
fits are consistent with the jackknife fits and  have reasonable
confidence levels. To get to the physical kaon I linearly extrapolated
my results with a jacknife; there is no sign of discernable curvature
 in my data.

I calculated the renormalization factors between the lattice- 
and continuum-regulated (NDR) matrix elements using one loop perturbation
theory. At $a$ (lattice spacing) $\times \mu$ (continuum regularization
point) $=1$, a conversion factor is $Z = 1 + (\alpha_s(q^*)/4\pi) z$.
As one might expect from related work\cite{Bernard:1999kc}, the HYP
link pushes the constant $z$ close to zero.
The cost is that $q^*$ (defined a la Lepage-Mackenzie\cite{ref:LM})
 can move to
a small value, but sensible values of $q^*$ are given by the higher-order
prescription of \cite{Hornbostel:2000ey}.

For $B_K$, the operator $O_+$ has a matching factor $Z_+$ and for the
overlap action used here, its
 parameter $z=-4.0$ at $q^* a = 0.92$ for NDR. 
The entire conversion factor for $B_K$
from lattice $\beta=5.9$ to $\mu=2$ GeV is 
$Z_+/Z_A^2=0.99$. (Wilson-action kernel overlap actions
 have $z$'s which are an order of magnitude larger.)

How reliable is this number? I have not checked it directly (yet),
but perturbation theory
for the local axial vector current
can be tested with overlap actions by a comparison of the
vacuum-to-pseudoscalar meson  matrix elements of the axial vector and
pseudoscalar density. At $\beta=5.9$ I find $Z_A=0.97$ or 0.98
(depending on the choice of lowest-order or higher-order $q^*$,
and 0.97(1) nonperturbatively. Perturbation theory for the matching
coefficient fo the $\overline{MS}$ quark mass can also be compared
to the nonperturbative calculation of Ref. \cite{Hernandez:2001yn}.
This analysis gives $Z(\mu=2 \ \rm{GeV}, a) =1.10(3)$ at $r_0 m_{PS}=5$ and
1.14(11) at $r_0 m_{PS}=3$, as compared to the perturbative prediction of 0.95.

 My PRELIMINARY values for $B_K$
and for other relevant parameters (lattice spacings from rho mass,
decay constants, $\overline{MS} (\mu=2$ GeV)
quark masses)
are recorded in the Table. A combined error from fitting,
extrapolation, and lattice spacing (dominated by statistics) is shown.
My $B_K$ result is in reasonably good agreement with the staggered
JLQCD result\cite{Aoki:1997nr}
a bit higher than the CP-PACS\cite{AliKhan:2001wr}
 domain wall fermion result
and quite a bit higher than the RBC\cite{Blum:2001xb}
 domain wall fermion result.
It is also consistent with the Wilson-overlap results presented by
Lellouch\cite{LELLOUCH} at this meeting.
Of course, it is a linear extrapolation: that is a dangerous thing to do.
 If the allocation gods allow, I hope to push
to smaller quark masses (bracketing the necessary quark mass and
possibly revealing the chiral logarithm)
and collect more statistics (hopefully giving more respectable error bars).

This work was supported by the US Department of Energy.
I are grateful to S. Sharpe for suggesting this project and
C. Bernard and
T. Blum for helpful instruction.
 Simulations were performed on the Platinum cluster at NCSA.

\end{document}